\documentclass[10pt]{iopart}
\usepackage[utf8]{inputenc}
\usepackage[square,sort&compress,numbers]{natbib}
\expandafter\let\csname equation*\endcsname\relax
\expandafter\let\csname endequation*\endcsname\relax
\usepackage{amsmath,amsfonts,graphicx,mathrsfs,amssymb,multirow}
\usepackage[colorlinks=true]{hyperref}
\usepackage{outlines}
\usepackage{subcaption,booktabs}

\renewcommand{\thefootnote}{\arabic{footnote}}

\begin{document}

\title{A Classical Firewall Transformation as a Canonical Transformation}

\author{Nathaniel A. Strauss, Bernard F. Whiting }
\address{Department of Physics, University of Florida, Gainesville, Florida, USA}
\eads{\mailto{straussn@ufl.edu}, \mailto{bernard@phys.ufl.edu}}

\date{\today}

\begin{abstract}
The firewall transformation put forward by 't Hooft in recent years has made ambitious claims of solving the firewall problem and the black hole information paradox while maintaining unitary evolution. However, the theory has received limited attention from the community, especially in regards to its foundations in purely classical gravitational physics. This paper investigates the underlying assumptions of 't Hooft's firewall transformation before quantization. We find that the limiting procedure used by 't Hooft in order to obtain an identification of the quantum operators for ingoing and outgoing particles near a black hole is not consistent. We propose a correction, which involves a more relaxed approximation regime. In the new approximation regime, we find a new classical analog for the firewall transformation for spherical shells, which allows evolving the spherical shells' dynamics past their point of collision. In the classical theory, no firewall is removed, as both ingoing and outgoing matter is present on every spacelike hypersurface, and it does not appear that any firewalls will be removed after a canonical quantization.

\end{abstract}

\maketitle

\renewcommand{\thefootnote}{\roman{footnote}}
\section{Introduction}

The behavior of matter near black holes is nontrivial in both quantum and classical physics. In this paper, we investigate two non-trivialities and how they interact: (1) particle creation near the black hole event horizon as predicted by quantum field theory and (2) the gravitational interaction of particles in a black hole background as predicted by general relativity. We start by discussing particle creation near black holes, known as Hawking radiation, which stems from a fusion of general relativity and quantum field theory. In quantum field theory, it can be shown that differently accelerating observers will disagree on the nature, or even existence of observed particles.~\citep{unruh_1976} In other words, there are many possible vacuum states, which are tailored to different observational frames.~\citep{fulling_1973,davies_1975} Applying the idea to a black hole background, a freely falling observer and an observer hovering just outside the horizon are accelerating with respect to one another, so if the freely falling observer does not detect particles, the observer hovering above the horizon will detect a thermal distribution of particles, which is known as Hawking radiation.~\citep{hawking_1974} There already existed a problem of information loss in classical mechanics as matter that falls beyond the event horizon looses its identity.  From a quantum mechanical perspective, the existence of Hawking radiation severely exacerbates the issue.  This happens because the entanglement between the black hole and the Hawking radiation results in the initial pure state evolving into a mixed state.  Furthermore, because Hawking radiation eventually causes the black hole to completely evaporate, the matter entangled to the Hawking radiation will eventually cease to exist.~\citep{mathur_2009} One resolution to the problem is to use unitary time evolution to include the Hawking radiation in the far past of the spacetime.  However, in order for the Hawking radiation to escape the black hole, it must have had very high energy near the event horizon in the far past, leading to a divergent stress energy on the horizon.~\citep{almheiri_2013,mathur_2009} This divergent stress energy is known as the firewall problem, and is not generally considered to be a real physical phenomenon. Rather, most physicists believe the firewall will disappear in a complete theory.~\citep{almheiri_2013,mathur_2009}

A proposal to resolve the firewall problem that preserves unitarity time evolution has been put forward by 't Hooft~\citep{'thooft_2016, 'thooft_2016_2,'thooft_2016_3,'thooft_2018,'thooft_2021,'thooft_2022}, and has continued to be developed in recent years with some engagement from the community in the context of searching for the quantum microstates of black holes (e.g.~\citep{hogan_2020,zeng_2021,kwon_2022,slagter_2022,egorov_2022}). 't Hooft's proposal is intended to resolve the firewall problem and information problem by including, into the quantum theory, the gravitational interaction of particles entering and leaving the black hole. The basic idea of 't Hooft's framework is that as the outgoing Hawking particles interact with ingoing matter, they leave informational ``footprints'' on each other, from which one can recover information that is lost in the standard theory. One allows firewalls to appear in the quantum theory, but one can also transform away any firewall that appears using a so-called ``firewall transformation''.~\citep{'thooft_2016_2,'thooft_2016_3}  't Hooft's model only considers radially moving particles for simplicity and splits the particles into two categories: ingoing and outgoing. As high energy particles pass by each other near the event horizon, they interact gravitationally, causing a Shapiro time delay and an exchange of energy. Normally the Shapiro time delay is thought of as an effect on particles scattering off a large massive body, and the time delay is the difference between the times it takes for the particle to reach some distant point with and without passing by the massive object. In this context, however, the time delay is caused by ingoing and outgoing particles passing by each other, not by the black hole. For example, in the context of this paper, the Shapiro time delay for an outgoing particle is the difference between the times it takes for the particle to reach a point far from the black hole with and without passing by one or more ingoing particles.

't Hooft suggests that including both the time delay caused by the high energy ingoing and outgoing particles and also keeping the high-energy particles in the theory would be double-counting the effect of the high-energy particles, so we can formally ignore the high-energy particles after including their gravitational effect on the surrounding particles.~\citep{'thooft_2016_2} As a consequence of this approach, the firewall transformation can be written in the following form for radially moving particles:~\citep{'thooft_2018}
\begin{align}
    \hat{u}_\text{in}(\Omega_j)&=\sum_i f(\Omega_j,\Omega_i) \hat{p}_\text{out}(\Omega_i),\label{eq:firewall_thooft_1}\\
    \hat{u}_\text{out}(\Omega_j)&=\sum_if(\Omega_j,\Omega_i) \hat{p}_\text{in}(\Omega_i),\label{eq:firewall_thooft_2}
\end{align}
where $\hat{u}_\text{in/out}$ and $\hat{p}_\text{in/out}$ represent an ingoing or outgoing particle's quantum position and momentum operators (in Kruskal coordinates), respectively, $\Omega_j$ and $\Omega_i$ represent the angular position of the particles relative to the black hole, and $f(\Omega_j,\Omega_i)$ gives the shift in particle $j$'s Kruskal position operator due to particle $i$. For example, a firewall on the future event horizon caused by the divergent momentum of ingoing particles can be removed by including the Shapiro time delay on the Kruskal-coordinate position of outgoing Hawking radiation.

In the authors' view, it is unclear whether the quantum firewall transformation given by~\eqref{eq:firewall_thooft_1} or~\eqref{eq:firewall_thooft_2} transforms away firewalls physically or merely formally. In other words, it is possible that the firewall physically exists in the theory even after it formally disappears from the notation. In order to elucidate the issue, and to orient us toward quantization, we take a close look at 
the classical analogs of~\eqref{eq:firewall_thooft_1} and~\eqref{eq:firewall_thooft_2} in a Hamiltonian framework. In order to follow 't Hooft's procedure as closely as possible, we specialize the Hamiltonian framework to Kruskal coordinates. For simplicity, our toy model will consist of two spherical shells of null matter, one ingoing and one outgoing.

The structure of the paper is as follows. In Section~\ref{sec:background}, we summarize the Shapiro time delay effect for two spherical shells of null matter passing by each other in a Schwarzschild background. In Section~\ref{sec:orders_of_magnitude}, we precisely describe the approximations necessary to obtain the classical analogs of~\eqref{eq:firewall_thooft_1} and~\eqref{eq:firewall_thooft_2}, even though the outcome is difficult to reconcile with 't Hooft's procedure.  In Section~\ref{sec:canonical}, we introduce the canonical theory for the two shells of null matter. In Section~\ref{sec:canonical_transformation}, we show that we can account for the time delay via a canonical transformation when we measure the time delay between the past of both shells and the future of both shells. Finally, in Section~\ref{sec:discussion}, we summarize our results and suggest further work is required in order to better understand to what degree 't Hooft's quantum firewall transformation may solve the firewall problem.

\section{The Shapiro Time Delay and the Firewall Transformation}
\label{sec:background}

The firewall transformation 't Hooft has proposed is motivated by including the gravitational shockwave of ingoing and outgoing particles passing by each other, which has been considered a Shapiro time delay effect.~\citep{'thooft_2021,'thooft_2022} In this section we will give an overview of the general relativity needed to understand the gravitational shockwave in a Schwarzschild background in the simplified case of radially moving spherical shells. We make use of results derived via the ADM Hamiltonian formalism, which is oriented towards canonical quantization.~\citep{louko_1998} The simplification to spherical shells removes all angular dependence to the resulting shockwave that point particles would create, but this angular dependence does not affect the larger context of the degree to which 't Hooft's firewall transformation solves the firewall problem.

\begin{figure}
    \centering
    \includegraphics{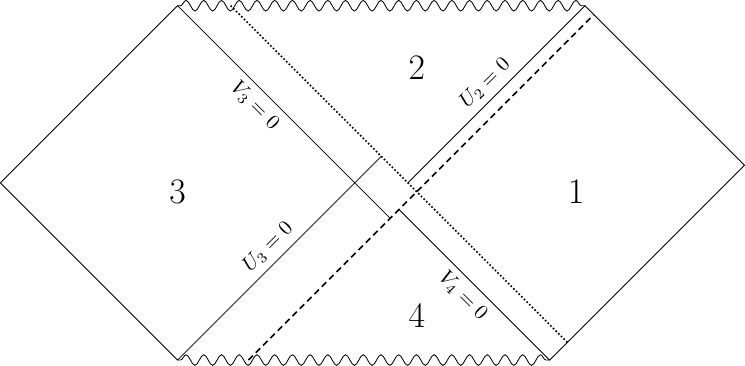}
    \caption{A Penrose diagram for a Schwarzschild black hole spacetime including two spherical shells of null matter, one ingoing and one outgoing. The ingoing shell is the dotted line and the outgoing shell is the dashed line. Lines of constant $U$ run from bottom left to top right and lines of constant $V$ run from bottom right to top left, both at an angle of 45 degrees. The diagram is partitioned into four labeled regions, separated by the two shells, which collide exterior to the event horizon. Note that region 1 lies entirely within the exterior of the black hole, while regions 2, 3, and 4 each contain part of the exterior and part of the interior.  Each of the four regions is installed with its own set of Kruskal coordinates $(U_i,V_i)$ and Schwarzschild mass $M_i$, and the visible coordinate axes are labeled in each region. Because the event horizon of the black hole moves when matter crosses it, the $U$ coordinate shifts when crossing the ingoing shell and the $V$ coordinate shifts when crossing the outgoing shell. }
    \label{fig:penrose_diagram}
\end{figure}

We begin with the Schwarzschild spacetime and insert two radially moving, spherical shells, as shown in Figure~\ref{fig:penrose_diagram}. In each of the four regions separated by the shells, the spacetime is given by the metric equation
\begin{align}
    \rmd s^2=-\Big(1-\frac{2M_i}{R}\Big) \rmd T_i^2 + \Big(1-\frac{2M_i}{R}\Big)^{-1} \rmd R^2 +R^2 \rmd\Omega^2
\end{align}
in ordinary Schwarzschild coordinates, where
\begin{align}
    \rmd\Omega^2&=\rmd\theta^2+\sin^2 \theta \rmd\phi^2,
\end{align}
and $M_i$ is the Schwarzschild mass in region $i$ of Figure~\ref{fig:penrose_diagram}. The metric can also be written in the Kruskal coordinates
\begin{align}
    \rmd s^2 = 2g_{U_iV_i}\rmd U_i \rmd V_i+ R^2 \rmd\Omega^2,
\end{align}
where
\begin{align}
    g_{U_iV_i}&=8M^2_i\frac{1-2M_i/R}{U_iV_i}=\frac{16 M^3_i}{R}\rme^{-R/2M_i}.
\end{align}
From this point forward, we suppress the indices $i$ when the expression applies to each region of Figure~\ref{fig:penrose_diagram} separately. For example, the Kruskal coordinates are related to the Schwarzschild coordinates implicitly by
\begin{align}
    UV&=\Big(\frac{R}{2M}-1\Big)\rme^{R/2M},\label{eq:uv_to_r}\\
    V/U&=\operatorname{sign}\Big(\frac{R}{2M}-1\Big)\rme^{T/2M}\label{eq:uv_to_t}
\end{align}
in each region separately.

Null particles traveling radially travel along lines of constant $U$ if they are outgoing and along lines of constant $V$ if they are ingoing, as will be shown in Section~\ref{sec:canonical}. 
It has long been known that the gravitational effect of a radially moving particle is an angular-dependent shift in one of the Kruskal coordinates $U$ or $V$ when matching the Kruskal coordinates across the trajectory of the particle.~\citep{dray_1985} In the case of a spherical shell of particles, the shift in the Kruskal coordinate of the shell depends only on the radius at the point of intersection, being angularly independent, and can be found via the definitions~\eqref{eq:uv_to_r} and~\eqref{eq:uv_to_t}. Let the total conserved energies of the ingoing and outgoing shells be
\begin{align}
    E_\text{in}&=M_1-M_4,\\
    E_\text{out}&=M_4-M_3
\end{align}
as measured in region 4 and
\begin{align}
    \tilde{E}_\text{in}&=M_2-M_3,\\
    \tilde{E}_\text{out}&=M_1-M_2
\end{align}
as measured in region 2. Defined this way, all $E$'s are positive.

By direct calculation using~\eqref{eq:uv_to_r}, we can find the matching conditions for the Kruskal coordinates across the shells to first order in the energies:
\begin{align}
    U_{1,\text{in}}-U_{4,\text{in}} &\approx -\frac{\rme^{R/2 M_4} R^2 E_\text{in}}{4 M^3_4 V_{4,\text{in}}},\label{eq:delta_U_1}\\
    U_{2,\text{in}}-U_{3,\text{in}} &\approx -\frac{\rme^{R/2 M_3} R^2 \tilde{E}_\text{in}}{4 M^3_3 V_{3,\text{in}}},\label{eq:delta_U_2}\\
    V_{1,\text{out}}-V_{2,\text{out}}&\approx -\frac{\rme^{R/2 M_2} R^2 \tilde{E}_\text{out}}{4 M^3_2 U_{2,\text{out}}},\label{eq:delta_V_1}\\
    V_{4,\text{out}}-V_{3,\text{out}}&\approx -\frac{\rme^{R/2 M_3} R^2 E_\text{out}}{4 M^3_3 U_{3,\text{out}}},\label{eq:delta_V_2},
\end{align}
where $U_{i,\text{in/out}}$ and $V_{i,\text{in/out}}$ are the Kruskal coordinates evaluated along the ingoing or outgoing shell, respectively, as measured in region $i$. Note that the shift in the Kruskal coordinate depends on the radius at which one matches the coordinates on each side of the shell. For example, a line of constant $U$ in region 4 will match to a different constant value of $U$ in region 1, depending on the radius at which that line impacts the ingoing shell, which travels along a line of constant $V$. Also, we make a gauge choice in our Schwarzschild time coordinate such that the Kruskal coordinate along the trajectory of each shell does not change when crossing that shell:
\begin{align}
    U_{1,\text{out}}&=U_{2,\text{out}},\label{eq:u_same_1}\\
    U_{4,\text{out}}&=U_{3,\text{out}},\label{eq:u_same_2}\\
    V_{1,\text{in}}&=V_{4,\text{in}},\label{eq:v_same_1}\\
    V_{2,\text{in}}&=V_{3,\text{in}}.\label{eq:v_same_2}
\end{align}
This above four gauge choices are not simultaneously consistent in the general theory, but are consistent to first order in the energies, the approximation regime in which we work. The result is a change in the Schwarzschild time coordinate (and not the Schwarzschild radial coordinate) when crossing the shell, which we call a Shapiro time delay.

When matching the coordinates across all four regions in Figure~\ref{fig:penrose_diagram}, a consistency condition emerges called the Dray-'t Hooft-Redmount formula~\citep{dray_1985,Redmount_1985}, which can be written
\begin{align}
    (R_0-2M_1)(R_0-2M_3)&=(R_0-2M_2)(R_0-2M_4),\label{eq:mass_condition}
\end{align}
where $R_0$ is defined by
\begin{align}
    U_{i,\text{out}}V_{i,\text{in}}&=\Big(\frac{R_0}{2M_i}-1\Big)\rme^{R_0/2M_i}
\end{align}
and represents the Schwarzschild radial coordinate where the two shells meet, and is assumed to lie in the exterior of all four Schwarzschild regions.  By rearranging~\eqref{eq:mass_condition}, we can show separately
\begin{align}
    \tilde{E}_\text{in}&=E_\text{in}+\frac{E_\text{in}E_\text{out}}{R_0-2M_4},\label{eq:energy_change_in}\\
    \tilde{E}_\text{out}&=E_\text{out}-\frac{E_\text{in}E_\text{out}}{R_0-2M_4}.\label{eq:energy_change_out}
\end{align}
We can thus think of the Dray-'t Hooft-Redmount formula as describing an exchange of energy between the two shells at the collision. Also, the conserved energy of the shells on the interior of the shells is larger because the Schwarzshchild mass is smaller there, which, roughly speaking, increases the shells' gravitational potential energy. We see that the outgoing shell loses energy after the shells meet, and the ingoing shell gains the same amount of energy after the shells meet. Nevertheless, this exchange of energy is second order in $E_\text{in}$ and $E_\text{out}$, so we neglect it for the remainder of the paper (see Section~\ref{sec:orders_of_magnitude}).  Note however that, unlike~\eqref{eq:delta_U_1} through~\eqref{eq:delta_V_2}, the results~\eqref{eq:mass_condition} through~\eqref{eq:energy_change_out} are all actually exact.

We have made a gauge choice in our Schwarzschild time coordinate such that the Kruskal coordinate along the trajectory of each shell does not change when crossing that shell. However, that Kruskal coordinate of each shell will change when the two shells meet. Also, the shift in the Kruskal coordinate approaches a finite value at the horizon of the black hole, corresponding to the shift in location of the event horizon due to the energy of the shell. In this paper we focus on the situation of the two shells colliding very close to the Schwarzschild horizon, i.e. $U_{i,\text{out}} V_{i,\text{out}}\ll 1$ and $r_0/2M-1\ll 1$. In this regime we can approximate the shifts in the constant Kruskal coordinates of the shells from~\eqref{eq:delta_U_1} through~\eqref{eq:delta_V_2} and~\eqref{eq:u_same_1} through~\eqref{eq:v_same_2}:
\begin{align}
    U_{1,\text{out}}-U_{3,\text{out}} &\approx -\frac{\rme  E_\text{in}}{ M_3 V_{3,\text{in}}},\label{eq:intermediate_firewall_1}\\
    U_{2,\text{out}}-U_{4,\text{out}} &\approx -\frac{\rme  E_\text{in}}{ M_4 V_{4,\text{in}}},\label{eq:intermediate_firewall_2}\\
    V_{1,\text{in}}-V_{3,\text{in}}&\approx -\frac{\rme  \tilde{E}_\text{out}}{ M_3 U_{3,\text{out}}},\label{eq:intermediate_firewall_3}\\
    V_{2,\text{in}}-V_{4,\text{in}}&\approx \frac{\rme  \tilde{E}_\text{out}}{ M_4 U_{4,\text{out}}}.\label{eq:intermediate_firewall_4}
\end{align}
These equations give the shifts in the constant Kruskal coordinate for each shell when the shells meet near the event horizon. After the shells meet, they continue to move along their new constant Kruskal coordinates. Because we are working to first order in $E_\text{in}$, $E_\text{out}$, $U_{i,\text{in}}$, and $V_{i,\text{out}}$, the right hand side of~\eqref{eq:intermediate_firewall_1} through~\eqref{eq:intermediate_firewall_4} will be the same on both sides of the shell, up to higher order terms (see Section~\ref{sec:orders_of_magnitude}). At this point, we've found that the change in Kruskal coordinate of the outgoing shell is proportional to the energy of the ingoing shell and vice versa, which forms the basis of a classical understanding for the firewall transformation.

\section{The approximation scheme}
\label{sec:orders_of_magnitude}
In obtaining the classical expressions~\eqref{eq:intermediate_firewall_1} through~\eqref{eq:intermediate_firewall_4}, which form the basis for the firewall transformation, we have made several approximations, which we will make more precise here. Firstly, we assume that the two shells meet near the event horizon of the black hole:
\begin{align}
    U_\text{out}&\sim \varepsilon_\text{out},\\
    V_\text{in}&\sim \varepsilon_\text{in},\label{eq:v_in_order}\\
    |U_\text{out}V_\text{in}|&=\Big|\frac{R_0}{2M}-1\Big|\rme^{R_0/2M}\sim \varepsilon_\text{out}\varepsilon_\text{in},
\end{align}
where the tilde signifies ``is of the order'' and $\varepsilon_\text{out},\varepsilon_\text{in}\ll 1$. We also neglect the change in energies of the shells across the shells given by~\eqref{eq:energy_change_in} and~\eqref{eq:energy_change_out}, imposing
\begin{align}
    \frac{2E_\text{out}}{R_0-2M}&\sim\delta_\text{out},\\
    \frac{2E_\text{in}}{R_0-2M}&\sim\delta_\text{in},
\end{align}
where $\delta_\text{in/out}\ll 1$. The above expressions imply
\begin{align}
    E_\text{in}&\sim \delta_\text{in}\varepsilon_\text{in} \varepsilon_\text{out},\label{eq:e_in_order}\\
    E_\text{out}&\sim \delta_\text{out} \varepsilon_\text{in}\varepsilon_\text{out},\\
    \tilde{E}_\text{in}-E_\text{in}&= -(\tilde{E}_\text{out}-E_\text{out})=\frac{2E_\text{in}E_\text{out}}{R_0-2M}\sim\delta_\text{in}\delta_\text{out} \varepsilon_\text{in}\varepsilon_\text{out}.
\end{align}
Therefore, we neglect terms of order $\delta_\text{in}\delta_\text{out} \varepsilon_\text{in}\varepsilon_\text{out}$. Now, using the notation of this section, we can show precisely to what order~\eqref{eq:intermediate_firewall_1} through~\eqref{eq:intermediate_firewall_4} are accurate. For example,
\begin{align}
    U_{1,\text{out}}-U_{3,\text{out}} &= -\frac{\rme^{R_0/2 M_4} R^2_0 E_\text{in}}{4 M^3 V_{3,\text{in}}}+O(\delta_\text{in}^2\varepsilon_\text{in} \varepsilon_\text{out}^2)\\
    &=-\frac{\rme  E_\text{in}}{M V_{3,\text{in}}}+O(\delta_\text{in}\varepsilon_\text{in} \varepsilon_\text{out}^2),\\
    V_{1,\text{in}}-V_{3,\text{in}}&= -\frac{\rme^{R_0/2 M_2} R^2_0 E_\text{out}}{4 M^3 U_{3,\text{out}}}+O(\delta_\text{out}^2 \varepsilon_\text{in}^2\varepsilon_\text{out})\\
    &=-\frac{\rme  E_\text{out}}{M U_{3,\text{out}}}+O(\delta_\text{out} \varepsilon_\text{in}^2\varepsilon_\text{out}).
\end{align}
Meanwhile, we can calculate directly
\begin{align}
    \frac{\rme  E_\text{in}}{M V_{3,\text{in}}}&\sim \delta_\text{in}\varepsilon_\text{out} ,\label{eq:delta_u_order} \\
    \frac{\rme  E_\text{out}}{M U_{3,\text{out}}}&\sim \delta_\text{out} \varepsilon_\text{in}.
\end{align}
Therefore, in order to obtain~\eqref{eq:intermediate_firewall_1} through~\eqref{eq:intermediate_firewall_4}, we work to third order in the regime $\varepsilon_\text{in} , \delta_\text{in} ,\varepsilon_\text{out}, \delta_\text{out} \ll 1$.

We now compare~\eqref{eq:intermediate_firewall_1} through~\eqref{eq:intermediate_firewall_4} to 't Hooft's firewall transformation in~\eqref{eq:firewall_thooft_1} and~\eqref{eq:firewall_thooft_2}. In 't Hooft's expressions, the initial Kruskal coordinate of each particle is taken to zero, i.e. the simultaneous limits $\varepsilon_\text{in}\rightarrow 0$ and $\varepsilon_\text{out}\rightarrow 0$. We see here that taking both limits also results in the change in both Kruskal coordinates  going to zero, i.e. $\delta_\text{in}\varepsilon_\text{out}\rightarrow 0$ and $\delta_\text{out} \varepsilon_\text{in}\rightarrow 0$. Thus, it is not consistent to take both initial Kruskal coordinates to zero while also keeping a non-zero shift in both Kruskal coordinates after crossing the shell. However, there is no problem taking exactly one of the limits $\varepsilon_\text{in}\rightarrow 0$ or $\varepsilon_\text{out}\rightarrow 0$, while keeping a non-zero shift in the appropriate coordinate. For example, taking the limit $\varepsilon_\text{in}\rightarrow 0$ means taking the simultaneous limits $V_{3,\text{in}}\rightarrow 0$ (seen from~\eqref{eq:v_in_order}) and $E_\text{in}\rightarrow 0$ (seen from~\eqref{eq:e_in_order}). We can take this limit such that $e E_\text{in}/MV_{3,\text{in}}$ remains of order $\delta_\text{in}\varepsilon_\text{out}$, as given by~\eqref{eq:delta_u_order}. The result is $V_{1,\text{in}}=V_{3,\text{in}}=0$, whereas $U_{1,\text{out}}-U_{3,\text{out}}\sim \delta_\text{in} \varepsilon_\text{out}$. To prevent this issue from obscuring any further results, we refrain from taking either of the limits $\varepsilon_\text{in}\rightarrow 0$ or $\varepsilon_\text{out}\rightarrow 0$ for the remainder of the paper.

\section{The Canonical Formalism for Spherical Shells in a Schwarzschild Background}
\label{sec:canonical}

We begin with the Hamiltonian formulation for a single shell.  In~\citep{louko_1998}, Louko et al. investigated the effect of a single spherical shell of null matter in a Schwarzschild background in great detail in the ADM formalism. In~\ref{sec:appendix_adm} we apply their results to the Kruskal coordinates, and find two expressions for the Hamiltonian and canonical momentum, with Schwarzschild time as the foliation time coodinate and one of the two Kruskal coordinates as the foliation radial coordinate (see~\eqref{eq:VHamiltonian.A} and~\eqref{eq:UHamiltonian.A}):
\begin{align}
    \mathcal{H}_U &= \frac{1}{2M}\frac{\eta_U \varepsilon_U-1}{2}p_{U} U,\\
    \mathcal{H}_V &=  \frac{1}{2M}\frac{\eta_V \varepsilon_V+1}{2}p_{V} V,
\end{align}
where
\begin{align}
    \varepsilon_U&=\operatorname{sign}U,\\
    \varepsilon_V&=\operatorname{sign}V,\\
    \eta_U&=\operatorname{sign}p_U,\\
    \eta_V&=\operatorname{sign}p_V.
\end{align}
Here, $p_V$ is the momentum conjugate to the Kruskal coordinate $V$ and $p_U$ is the momentum conjugate to the Kruskal coordinate $U$. 

The two Hamiltonians are related by a canonical coordinate transformation generated by a type two generating function. For example, when transforming from $V$ to $U$, the generating function
\begin{align}
    G_2 = \operatorname{sign}(R-2M)\rme^{-T/2M} p_U V
\end{align}
results in the transformation equations
\begin{align}
    U=\operatorname{sign}\Big(\frac{R}{2M}-1\Big)\rme^{-T/2M}V,\label{eq:canonical_coordinates}\\ p_U = \operatorname{sign}\Big(\frac{R}{2M}-1\Big)\rme^{T/2M} p_V,\label{eq:canonical_momenta}
\end{align}
which is consistent with~\eqref{eq:uv_to_t}. Furthermore, in transforming from $V$ to $U$ using $G_2$ as the generating function, the Hamiltonian changes from $\mathcal{H}_V$ to $\mathcal{H}_U$. 

Via canonical coordinate transformations between the canonical variables $(U,p_U)$ and $(V,p_V)$, we can obtain a meaningful Hamiltonian and meaningful equations of motion using either Kruskal coordinate, regardless of whether the shell is ingoing or outgoing. In the remainder of this paper, we are only concerned with the dynamics in the exterior of the Schwarzschild horizon, so we take $\varepsilon_U=\varepsilon_V=1$. We also wish to clearly distinguish the ingoing shell from the outgoing shell, regardless of canonical variables, either $(U,p_U)$ or $(V,p_V)$, we use to describe that shell. We use the subscript ``in'' to signify the quantity for the ingoing shell, where $\eta_U=\eta_V=-1$. Likewise, we use the subscript ``out'' to signify the quantity for the outgoing shell, where $\eta_U=\eta_V=+1$. For example, the Hamiltonians are
\begin{align}
    \mathcal{H}_{U\text{in}} &= -\frac{1}{2M}p_{U\text{in}} U_\text{in},\\
    \mathcal{H}_{V\text{in}} &=  0,\\
    \mathcal{H}_{U\text{out}} &= 0,\\
    \mathcal{H}_{V\text{out}} &=  \frac{1}{2M}p_{V\text{out}} V_\text{out},
\end{align}
where for each shell, we use either $\mathcal{H}_U$ or $\mathcal{H}_V$, and we can switch between them via a canonical coordinate transformation. Notice in one set of variables for each shell the Hamiltonian is zero, which means in those variables both the canonical coordinate and the momentum are constants of motion. This justifies our earlier assertion that ingoing shells travel along constant $V$ and outgoing shells travel along constant $U$.

By coupling the action for the shell to the gravitational action, Louko et al. found a relationship between shell's momenta and the radial derivative of the foliation coordinates.~\citep{louko_1998} Applying those results to the Kruskal coordinates, we find the following equations of motion for the momenta:
\begin{align}
    p_{U\text{in}}&=-\frac{4ME_\text{in}}{V_{\text{in}}}e^{T/2M},\label{eq:momenta_1}\\
    p_{V\text{in}}&=-\frac{4ME_\text{in}}{V_{\text{in}}},\\
    p_{U\text{out}}&=\frac{4ME_\text{out}}{U_{\text{out}}},\\
    p_{V\text{out}}&=\frac{4ME_\text{out}}{U_{\text{out}}}e^{-T/2M}.\label{eq:momenta_4}
\end{align}
Expressions~\eqref{eq:momenta_1} through~\eqref{eq:momenta_4} are defined in each region of Figure~\ref{fig:penrose_diagram} separately, though the change in each symbol on the right hand side is at most second order in $\delta_\text{in}$ or $\delta_\text{out}$, so in our approximation regime, we can assume the momenta of the shells are unambiguous all along each shell, and at the point of intersection. We have a physically meaningful canonical momentum in both sets of Kruskal variables, regardless of whether the particle is ingoing or outgoing, and regardless of in which region of Figure~\ref{fig:penrose_diagram} we observe the shell.

Returning to the case when two intersecting shells are present throughout the spacetime, we can now rewrite~\eqref{eq:intermediate_firewall_1} through~\eqref{eq:intermediate_firewall_4} in terms of the canonical variables:
\begin{align}
    U_{1,\text{out}}-U_{3,\text{out}} &= \frac{e}{4M^2_3}p_{3,V\text{in}},\label{eq:firewall_space_1}\\
    V_{1,\text{in}}-V_{3,\text{in}}&= -\frac{e}{4M^2_3}p_{3,U\text{out}},
\end{align}
or equivalently
\begin{align}
    U_{2,\text{out}}-U_{4,\text{out}} &= \frac{e}{4M^2_4}p_{4,V\text{in}},\\
    V_{2,\text{in}}-V_{4,\text{in}}&= \frac{e}{4M^2_4}p_{4,U\text{out}}\label{eq:firewall_time_2}.
\end{align}
In one set the two compared regions are spacelike separated and in the other they are timelike separated. In order to streamline the notation and concisely summarize the matching of the Kruskal coordinates of the shells in  the different regions, we write
\begin{align}
    U_{\text{out}}'-U_{\text{out}} &= \sigma_\text{out}\frac{e}{4M^2}p_{V\text{in}},\label{eq:general_firewall_1}\\
    V_{\text{in}}'-V_{\text{in}}&=  \sigma_\text{in} \frac{e}{4M^2}p_{U\text{out}}\label{eq:general_firewall_2},
\end{align}
where the values of $\sigma_\text{out}$ and $\sigma_\text{in}$ are given in Table~\ref{tab:sigmas}, depending on the regions in which $U_\text{out}$, $U_\text{out}'$, $V_\text{in}$, and $V_\text{in}'$ are defined. We consider these two equations to be our precise classical analogs of 't Hooft's firewall transformation equations,~\eqref{eq:firewall_thooft_1} and~\eqref{eq:firewall_thooft_2}. The goal for the remainder of the paper is to include~\eqref{eq:general_firewall_1} and~\eqref{eq:general_firewall_2} in the canonical theory, oriented towards canonical quantization.
\begin{table}[t]
    \begin{subtable}{0.48\textwidth}
    \centering
    \begin{tabular}{cc|cccc}
    \multicolumn{6}{c}{\large $\sigma_\text{out}$}\vspace{0.2cm}\\ 
        &\multicolumn{1}{c}{} &\multicolumn{4}{c}{$U_\text{out}'$ region}\\
        &  & 1 & 2 & 3 & 4\\ \cline{2-6}
         \multirow{4}{*}{\rotatebox[origin=c]{90}{$U_\text{out}${\,}region}}& 1 & 0 & 0 & $-1$ & $-1$ \\
         & 2 & 0 & 0 & $-1$ & $-1$ \\
         & 3 & $1$ & $1$ & 0 & 0\\
         & 4 & $1$ & $1$ & 0 & 0
    \end{tabular}
    \caption{}
    \label{tab:sigma_out}
    \end{subtable}
    \begin{subtable}{0.48\textwidth}
    \centering
    \begin{tabular}{cc|cccc}
    \multicolumn{6}{c}{\large $\sigma_\text{in}$}\vspace{0.2cm}\\ 
        &\multicolumn{1}{c}{} &\multicolumn{4}{c}{$V_\text{in}'$ region}\\
        &  & 1 & 2 & 3 & 4\\ \cline{2-6}
         \multirow{4}{*}{\rotatebox[origin=c]{90}{$V_\text{in}$ region}}& 1 & 0 & 1 & 1 & 0\\
         & 2 & $-1$ & 0 & 0 & $-1$\\
         & 3 & $-1$ & 0 & 0 & $-1$\\
         & 4 & 0 & 1 & 1 & 0
    \end{tabular}
    \caption{}
    \label{tab:sigma_in}
    \end{subtable}
    \caption{These two tables give the values of $\sigma_\text{out}$ and $\sigma_\text{in}$ in~\eqref{eq:general_firewall_1} and~\eqref{eq:general_firewall_2} depending on the regions in which $U_\text{out}$, $U_\text{out}'$, $V_\text{in}$, and $V_\text{in}'$ are defined. Table~\ref{tab:sigma_out} gives the value of $\sigma_\text{out}$ and Table~\ref{tab:sigma_in} gives the value of $\sigma_\text{in}$.}
    \label{tab:sigmas}
\end{table}

\section{The Firewall Transformation as a Canonical transformation}
\label{sec:canonical_transformation}
The firewall transformation as written in~\eqref{eq:firewall_thooft_1} and~\eqref{eq:firewall_thooft_2} is a simple change of basis in the Hilbert space for the wavefunction describing the quantum particles. Perhaps the most straightforward way to obtain 't Hooft's firewall transformation after canonical quantization is to demand the relations~\eqref{eq:general_firewall_1} and~\eqref{eq:general_firewall_2} hold as canonical coordinate transformations. Consider a canonical Hamiltonian containing one ingoing shell and one outgoing shell:
\begin{align}
    \mathcal{H}_c &= \frac{p_{V\text{out}} V_\text{out}}{2M}-\frac{p_{U\text{in}} U_\text{in}}{2M},\label{eq:canonical_hamiltonian}
\end{align}
where the Hamiltonian is defined in each region of Figure~\ref{fig:penrose_diagram} separately. In order to use our expressions for the firewall transformation~\eqref{eq:general_firewall_1} and~\eqref{eq:general_firewall_2}, we need to first perform a canonical coordinate transformation to the canonical variables from $(U_\text{in},p_{U\text{in}})$ and $(V_\text{out},p_{V\text{out}})$ to $(U_\text{out},p_{U\text{out}})$ and $(V_\text{in},p_{V\text{in}})$. This renders the Hamiltonian equal to zero, since the canonical coordinates and momenta are constants of the motion, but they are still physically meaningful via~\eqref{eq:momenta_1} through~\eqref{eq:momenta_4}. Suppose now we transform to new canonical variables
\begin{align}
    U_{\text{out}}'(U_\text{out},p_{V\text{in}})&=U_{\text{out}} + \sigma_\text{out}\frac{e}{4M^2}p_{V\text{in}},\label{eq:can_trans_U}\\ V_{\text{in}}'(V_\text{in},p_{U\text{out}})&=V_{\text{in}}+  \sigma_\text{in} \frac{e}{4M^2}p_{U\text{out}},\\
    p_{U\text{out}}'&=p_{U\text{out}},\\
    p_{V\text{in}}'&=p_{V\text{in}},\label{eq:can_trans_pV}
\end{align}
which are consistent with~\eqref{eq:general_firewall_1} and~\eqref{eq:general_firewall_2}. We can check if the transformation is canonical by calculating the Poisson brackets:
\begin{align}
    \{U_\text{out}',p_{U\text{out}}'\}&=1,\\
    \{V_\text{in}',p_{V\text{in}}'\}&=1,\\
    \{U_\text{out}',p_{V\text{in}}'\}&=0,\\
    \{V_\text{in}',p_{U\text{out}}'\}&=0,\\
    \{p_{U\text{out}}',p_{V\text{in}}'\}&=0,\\
    \{U_\text{out}',V_\text{in}'\}&= \frac{e}{4M^2}(\sigma_\text{in}-\sigma_\text{out}),
\end{align}
where $\{,\}$ signifies the Poisson bracket in the old variables $(U_\text{out},p_{U\text{out}})$ and $(V_\text{in},p_{V\text{in}})$. Thus, the transformation is only canonical when $\sigma_\text{out}=\sigma_\text{in}$.

By examining Table~\ref{tab:sigmas}, we find two significant cases when $\sigma_\text{out}=\sigma_\text{in}$: (1) the primed variables are defined in region 2 and the unprimed variables are defined in region 4 and (2) the unprimed variables for the ingoing shell and the primed variables for the outgoing shell are defined in region 1, and the primed variables for the ingoing shell and the unprimed variables for the outgoing shell are defined in region 3. In both cases, the unprimed variables describe the shells before the collision, and the primed variables describe the shells after the collision. The other cases for which $\sigma_\text{out}=\sigma_\text{in}$ involve either switching the primed and unprimed regions in the above two cases or the trivial cases where the primed and unprimed regions are identical. Thus, we have found that we can interpret~\eqref{eq:general_firewall_1} and~\eqref{eq:general_firewall_2} as a canonical transformation between the canonical variables in the past of the point of collision and the future of the point of collision.

Thus, we see that the canonical transformations identified in cases (1) and (2) above define how to evolve the two null shells past the collision point without introducing the gravitational degrees of freedom into the canonical theory. The fact that the transformation is canonical only when evolving between the past and future of the shells suggests that canonical transformations between other regions are missing evolutionary data. The canonical transformation approach is also advantageous because it is straightforward to quantize, the transformation equations surviving as equalities between operators. However, before or after quantization, it does not appear that the canonical ``firewall transformation'' given by~\eqref{eq:general_firewall_1} and~\eqref{eq:general_firewall_2} removes any firewalls from the geometry, since both shells remain present in the theory on all spacelike hypersurfaces.

\section{Discussion}
\label{sec:discussion}

The firewall transformation, given by~\eqref{eq:firewall_thooft_1} and~\eqref{eq:firewall_thooft_2}~\citep{'thooft_2018}, is inspired by the purely classical Shapiro time delay that results from energetic bodies passing by each other and aims to resolve the firewall problem while preserving unitarity of the time-evolution in the quantum theory. However, it remains unclear to the present authors to what degree the firewall transformation actually solves the firewall problem. In order to elucidate the issue, we have proposed methods for including the Shapiro time delay into the classical theory, specialized to the case of radially moving spherical shells of null matter, and we have oriented our approach towards quantization by using the canonical ADM formalism. Our first result is from Section~\ref{sec:orders_of_magnitude}, in which we showed that taking the initial Kruskal coordinates of both the ingoing and outgoing shells to zero is inconsistent with keeping a nonzero shift in those coordinates. This result is not consistent with 't Hooft's procedure~\citep{'thooft_2018} of taking both limits to zero in the quantum theory, which obscures the physical meaning of 't Hooft's quantum firewall transformation. We have corrected the issue by taking only one or neither of the Kruskal coordinates of the ingoing or outgoing shells to zero. In Section~\ref{sec:canonical_transformation}, we showed that we can impose the Shapiro time delay between the canonical variables before and after the collision by demanding the canonical variables for the shells in each region are related by a canonical coordinate transformation, which we interpret as a way to evolve the shells beyond the point of collision. 

In the quantum case discussed by 't Hooft, the firewall transformation removes all ingoing particles or all outgoing particles at the horizon, which in turn removes the firewall that those particles create. However, in our classical analog, we have a Hamiltonian including two spherical shells at all times, even after performing a canonical transformation to include the Shapiro time delay. Exactly how one proceeds from the classical theory to the quantum firewall transformation has so far not been made explicit in the literature. Thus, either the quantum firewall transformation's removal of firewalls is a purely quantum effect, or the quantum firewall transformation merely obscures any firewalls by removing them formally without removing them physically. We leave to future work any investigation of the corresponding quantum theory for the classical approach we have outlined in this paper. Regardless of whether or not the quantum firewall transformation removes firewalls, it appears to preserve at least some information normally lost beyond the event horizon of the black holes in the observables for the outgoing particles, since the firewall transformation relates ingoing and outgoing observables to each other. Furthermore, if the firewall transformation is able to provide a unitary scattering matrix for particles interacting near a black hole, that would be a significant step forward.

We make one final remark concerning case (2) of Section~\ref{sec:canonical_transformation}.  When referring to the firewall transformation, 't Hooft has often spoken in terms of suppressing half of the physical variables.  Exactly this happens if we refer to physical degrees of freedom only in region 1 in the context of case (2), while suppressing altogether the physical degrees of freedom in region 3 (which could, nevertheless, be reconstructed by reference to the firewall transformation).  A feature that is not quite captured by our shell model is that every ingoing particle will interact with every outgoing particle.  It might have been the case that every ingoing particle lost energy during each interaction, and every outgoing particle gained energy as a result of each interaction.  Then we could imagine that, finally, all ingoing particles would end up with zero energy, and could be ignored anyway, while all outgoing particles would start out with zero energy, and hence could be ignored to begin with -- so that some of what 't Hooft seeks could perhaps be recaptured -- but none of that is what our calculations indicate. Instead, the ingoing particles all gain energy as they interact with each outgoing particle, and the outgoing particles all lose energy as they interact with each ingoing particle, so both sets of particles are maximally energetic as they cross their respective horizons, exactly where the firewalls would seem to exist.  It is precisely here that 't Hooft wishes all particles to be ignored.  However, to us, the choice to ignore all particles in region 3, where indeed they are most energetic, seems to be somewhat bewildering, and is a choice yet to be fully fathomed.


\section*{Acknowledgements}  
NAS acknowledges support from the Graduate Student Fellowship, the CLAS Dissertation Fellowship, and a teaching assistantship from the University of Florida.  BFW acknowledges support from NSF through grant PHY-1607323, and from the University of Florida.  Hospitality at the Observatoire de Meudon, the Institut d'Astrophysique de Paris, and at the Albert Einstein Institute in Potsdam throughout the course of this work is also gratefully acknowledged.
\vfill\eject

\bibliographystyle{iopart-num}
\bibliography{references}

\pagebreak 

\appendix

\section{}
\label{sec:appendix_adm}
In this appendix, we derive the Hamiltonians for spherical shells of null matter via the ADM canonical formalism for general relativity with spherical symmetry. The ADM formalism with spherical symmetry admits the following decomposition of the metric:~\citep{louko_1998}
\begin{align}
    \rmd s^2&=-N^2\rmd t^2+\Lambda^2(\rmd r+N^r \rmd t)^2 +R^2\rmd\Omega^2\nonumber\\
    &=-(N^2-\Lambda^2 N^r{}^2)\rmd t^2+2\Lambda^2 N^r \rmd t \rmd r + \Lambda^2 \rmd r^2+R^2 \rmd\Omega^2,\label{eq:adm_metric}
\end{align}
where $t$ and $r$ are the foliation coordinates, $N(t,r)$ is the lapse, $N^r(t,r)$ is the shift, and $\Lambda(t,r)$ and $R(t,r)$ are the canonical variables of the metric. The action for a spherical shell of matter is \begin{align}
    S&=\int \rmd t \mathcal{L}=-m\int \rmd t \sqrt{\hat{N}^2-\hat{\Lambda}^2(\dot{\mathfrak{r}}+\hat{N}^r)^2},\label{eq:particle_action}
\end{align} 
where $m$ is the rest mass of the shell, $\mathfrak{r}(t)$ is the canonical coordinate $r$ for the shell, a dot is a partial derivative with respect to $t$, and a hat signifies the quantity is evaluated on the shell. The momentum conjugate to $\mathfrak{r}$ is
\begin{align}
    \mathfrak{p}=\frac{\delta \mathcal{L}}{\delta \dot{\mathfrak{r}}}=\frac{m\hat{\Lambda}^2(\dot{\mathfrak{r}}+\hat{N}^r)}{\sqrt{\hat{N}^2-\hat{\Lambda}^2(\dot{\mathfrak{r}}+\hat{N}^r)^2}}.
\end{align}
Solving directly for the velocity,
\begin{align}
    \dot{\mathfrak{r}}=\eta\frac{\hat{N}/\hat{\Lambda}}{\sqrt{1+\hat{\Lambda}^2m^2/\mathfrak{p}^2}}-\hat{N}^r,
\end{align}
where $\eta\equiv \operatorname{sign}\mathfrak{p}$. Thus, in the null limit $m\rightarrow 0$, we have
\begin{align}
    \dot{\mathfrak{r}}=\eta\frac{N}{\Lambda}-N^r.
\end{align}
We can then construct the Hamiltonian:
\begin{align}
    \mathcal{H}&=\mathfrak{p}\dot{\mathfrak{r}}-\mathcal{L}\nonumber\\
    &=\mathfrak{p}\Big(\eta\frac{\hat{N}}{\hat{\Lambda}}-\hat{N}^r\Big).~\label{eq:adm_hamiltonian}
\end{align}

In this paper, we specialize the approach to Kruskal coordinates, given by the metric
\begin{align}
    \rmd s^2 = 2g_{UV}\rmd U \rmd V+ R^2 \rmd\Omega^2,
\end{align}
where
\begin{align}
    g_{UV}&=8M^2\frac{1-2M/R}{UV}=\frac{16 M^3}{R}\rme^{-R/2M}.
\end{align}
The Kruskal coordinates are related to the Schwarzschild coordinates implicitly by
\begin{align}
    UV&=\Big(\frac{R}{2M}-1\Big)\rme^{R/2M},\\
    V/U&=\operatorname{sign}\Big(\frac{R}{2M}-1\Big)\rme^{T/2M}.
\end{align}
In the construction of the shell Hamiltonians we utilize hybrid coordinate systems using the Schwarzschild time coordinate $T$ as the foliation time coordinate $t$ and either Kruskal coordinate $U$ or $V$ as the foliation radial coordinate $r$. The metrics are 
\begin{align}
    ds^2 &= 2e^{-T/2M} g_{UV}(dV^2-\tfrac{V}{2M}dTdV)+R^2 d\Omega^2\\
    &= 2e^{T/2M} g_{UV}(dU^2+\tfrac{U}{2M}dTdU)+R^2 d\Omega^2,
\end{align}
and comparing with~\eqref{eq:adm_metric}, we find the corresponding ADM variables:
\begin{align}
    N^2 &= 1-\frac{2M}{R},& \Lambda^2 &= 2g_{UV}e^{-T/2M},&  N^r &= -\frac{V}{4M},& r&=V,& t&=T
\end{align}
with $V$ as the radial coordinate and
\begin{align}
    N^2&=1-\frac{2M}{R},& \Lambda^2 &= 2g_{UV}e^{T/2M},& N^r &= \frac{U}{4M},& r&=U, & t&=T
\end{align}
with $U$ as the radial coordinate. Inserting the above expressions into~\eqref{eq:adm_hamiltonian}, we find the Hamiltonians specialized to the Kruskal coordinates with the Schwarzschild time coordinate:
\begin{align}
    \mathcal{H}_V&=\mathfrak{p}\Big(\eta\frac{\hat{N}}{\hat{\Lambda}}-\hat{N}^r\Big)\nonumber\\
    &=\frac{1}{2M}\frac{\eta \varepsilon+1}{2}\mathfrak{p} \hat{V},\label{eq:VHamiltonian.A}\\
    \mathcal{H}_U&=\mathfrak{p}\Big(\eta\frac{\hat{N}}{\hat{\Lambda}}-\hat{N}^r\Big)\nonumber\\
    &=\frac{1}{2M}\frac{\eta \varepsilon-1}{2}\mathfrak{p} \hat{U}\label{eq:UHamiltonian.A}
\end{align}
where $\varepsilon$ is the sign of $\hat{V}$ or $\hat{U}$, respectively, and $\mathfrak{r}$ is equal to $\hat{V}$ or $\hat{U}$, respectively.  Equations~\eqref{eq:VHamiltonian.A} and~\eqref{eq:UHamiltonian.A} represent our starting point in Section \ref{sec:canonical}.

\end{document}